\documentclass[twocolumn,showpacs,preprintnumbers,amsmath,amssymb]{revtex4}

\usepackage{graphicx}
\usepackage{dcolumn}
\usepackage{bm}

\begin{document}


\title{Fractal design for efficient brittle plates under gentle pressure loading}

\author{Robert S. Farr}
 \affiliation{Unilever R\&D, Olivier van Noortlaan 120, AT3133, Vlaardingen, The Netherlands}
 \email{robert.farr@unilever.com}

\date{\today}

\begin{abstract}
We consider a plate made from an isotropic but brittle elastic material, 
which is used to span a rigid aperture, across which a small pressure 
difference is applied. 
The problem we address is to find the structure which uses the least amount of 
material without breaking. Under a simple set of physical approximations and 
for a certain region of the pressure-brittleness parameter space, we find that 
a fractal structure in which 
the plate consists of thicker spars supporting thinner spars in an
hierarchical arrangement gives a design of high mechanical efficiency.
\end{abstract}


\maketitle

\section{Introduction}
In the construction of buildings and implements, optimising mechanical 
efficiency (minimising the amount of material used)
has been practiced since time immemorial. Striking examples may be 
seen in the flying buttresses and remarkably thin vaulted roofs 
of medieval European cathedrals.
The underlying principles for reducing the mass of masonry 
were developed much later however, starting with the concept of thrust 
lines \cite{Heyman}.

After Euler \cite{Euler} studied the buckling of struts, it was
recognised that compression members are in general considerably less 
efficient than tension members. This line of reasoning was pursued
in the twentieth century by researchers (e.g. \cite{Michell,Cox}) 
who also considered the question of couplings
required to attach different parts of engineering structures to one another.
Because of the scaling properties of Euler buckling and
the cost in material for couplings of
tension members, it is generally advantageous 
to have few compression members and many long tension members.
An efficient structure such as a space frame will
therefore typically resemble a tent in this regard; 
a pattern which is 
also seen in the anatomy of land animals where compressional
limb bones are sheathed in a finely divided web of tendons and musculature
\cite{Gordon}.

In more recent years, the main focus in structural optimization has 
switched to computational approaches. Various
classes of problem are considered: for example under a fixed set of
loads and using a fixed total volume of material, to minimise the 
compliance (energy storage) or the maximum stress in the structure.
For trusses, the ``ground structure method'' \cite{Bendsoe1} is typically
applied, where the initial structure is specified by a set of $n$
points, and
a spaceframe described by (a subset of) the complete graph $K_n$
which has these points as nodes.
The cross sections of the beams are then varied (potentially down 
to zero size).

For the optimization of solid components, the naive approach of drilling
holes to reduce weight has been brought to a high degree of refinement
with methods such as the ``SIMP'' scheme 
(for ``Solid Isotropic Microstructure with Penalization'') 
\cite{Bendsoe2,Eschenauer} where voxels of
the material are allowed to have grey-scale values during the optimization
process. These so-called topology optimization methods allow for any number
and shape of holes to emerge from the simulation, except that a minimum length
scale (larger than the voxel size) is imposed \cite{Bendsoe2}
because of the tendency
for optimal structures to contain many fine tension members \cite{Cox}
which are difficult to manufacture.

As noted above, the tendency towards fine subdivision of tension members
is well known. However a rather different possibility has emerged from
two widely different fields. This is the idea that fractals \cite{Mandelbrot},
which are well known to have interesting mechanical
properties when they occur in colloidal flocs \cite{Lin,Kantor,Buscall} 
and have recently been shown to give highly efficient transport networks
\cite{West} might also occur as optimum mechanical designs:

Firstly it has been noted that
trabecular bone has a fractal architecture, which has
been hypothesized to be related to mechanical efficiency \cite{Huiskes1}.
Unfortunately, more recent work indicates that this complex microstructure 
may not be the result of such a simple optimization \cite{Huiskes2}.

Secondly, in paleontology the complex suture patterns of ammonites have
been seen as an adaptation for greater strength \cite{Jacobs,Batt}; but
again more recent work suggests that fractal morphology is not correlated to
life at greater depth \cite{Oloriz}. Note however that this evidence 
is not conclusive, since
strength and efficiency are different quantities;
greater efficiency from fractal design may be demanded by lower availability
of minerals and (as a very tentative extrapolation from results in this paper) 
be possible only at shallower depth.

Given the tantalising possibilities and the complexity of the systems 
previously studied, it would be interesting to construct a problem in 
mechanical structure optimisation which is simple enough to analyse in depth, 
yet encompassing enough physics to be non-trivial. Careful analysis
might then suggest whether fractal design principles may indeed give highly 
efficient structural solutions.

\begin{figure}
\includegraphics[width=\columnwidth]{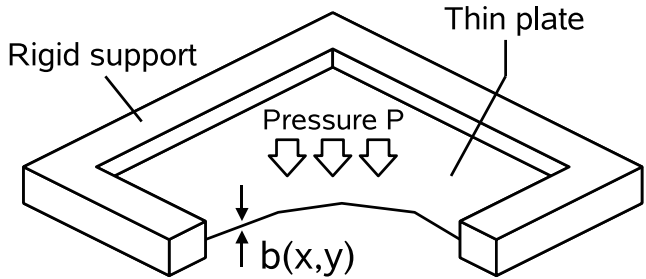}
\caption{\label{panel} 
Cutaway view of a thin pressure-supporting plate which 
spans a square aperture, and is clamped at the edges of this to a rigid support.  }
\end{figure}

To this end, we consider the following problem: suppose we have an aperture 
in a rigid structure, which we wish to span with a thin plate of an 
elastically isotropic but brittle material (Fig.~\ref{panel}). Furthermore, 
suppose that we want this opening to support a pressure difference, so that 
the spanning plate has a tendency to bow or to break. We aim to find the 
minimum amount of material required to support this pressure without breaking. 
We also impose the condition that the plate is unstressed in the absence of 
the pressure loading. Such a situation might arise in different contexts; 
for example in the manufacture of silicon membranes for membrane 
emulsification \cite{Joscelyne} or the construction of pressure vessels.

\section{Precise statement of the problem}
Consider a thin plate of elastic material with Young 
modulus $Y$ and Poisson ratio $\nu$.
We assume that when the plate is not subject to a pressure load, it 
has no quenched-in stresses, it lies in the $x-y$ plane, and has a 
possibly non-uniform thickness $b(x,y)$. 

The plate is attached to the boundary of a rigid aperture, 
so that even under load, the out-of-plane deflection and its gradient are 
zero at the boundary. We refer to these constraints 
as ``cantilever boundary conditions''. 

Let $L_0$ be a typical unit of length 
(for example of the same order of magnitude as the size of the aperture), 
then we assume that the plate is thin, in the sense that $b/L_0\ll 1$. 
Let us suppose that the plate is now subject to a small pressure 
difference $P$ between its 
two sides, and we define a non-dimensional pressure loading parameter $p=P/Y$.

In order to state the problem, we first need an expression for the elastic 
energy of the plate under such pressure loading. We are furthermore 
interested in
the case where the deflection of the plate may not be small compared
to its thickness, and we will therefore have to consider stretching
or shearing of the middle plane of the plate \cite{Timoshenko}.

Under these circumstances, a useful approximation is that of von Karman
\cite{vonKarman,Timoshenko}. A simple exposition may also be found in
Ref.~\cite{Lobkovsky}, from which we take (with a change of notation)
the results we need below.

Consider a region of the middle plane of the plate near a point 
which has co-ordinates before loading given by $(x_0,y_0,0)$. 
Under a load $p$, the plate will deform, after which we define the new 
co-ordinates of the point under consideration to be 
\[\left( x_0+u(x_0,y_0),y_0+v(x_0,y_0),w(x_0,y_0)\right),\]
where $u$, $v$ and $w$ are assumed to be small and we have allowed the 
possibility of out-of-plane deformations $w$.

The two-dimensional strain tensor of the middle layer after deformation
is given by ${\rm\bf e}$ with components
\begin{eqnarray}
e_{xx}=\frac{\partial u}{\partial x_0}+\frac{1}{2}
\left(\frac{\partial w}{\partial x_0}\right)^2
\nonumber \\
e_{xy}=\frac{1}{2}\frac{\partial u}{\partial y_0}+
\frac{1}{2}\frac{\partial v}{\partial x_0}+\frac{1}{2}
\left(\frac{\partial w}{\partial x_0}\right)
\left(\frac{\partial w}{\partial y_0}\right)
\nonumber \\
e_{yy}=\frac{\partial v}{\partial y_0}+\frac{1}{2}
\left(\frac{\partial w}{\partial y_0}\right)^2
\nonumber
.
\end{eqnarray}
For a uniformly thick plate (or a plate with sufficiently slowly
varying thickness \cite{Timoshenko2}) the equilibrium deformation of
the middle plane can be obtained by minimising the energy $U$ of 
Eq.~(\ref{U}) with suitable boundary conditions \cite{Lobkovsky}, where
\begin{equation}\label{U}
U=U_S+U_B-Y\int{dxdy(pw)}.
\end{equation} 
and the stretching and bending energies of the plate can be simply derived
from the corresponding expressions in Ref.~\cite{Lobkovsky} as:

\begin{equation}\label{US}
U_{S}=\int dxdy\frac{Yb}{2(1-\nu^2)}\left\{
\nu\left[ {\rm Tr}({\rm\bf e})\right]^2+(1-\nu){\rm Tr}({\rm\bf e}^2)\right\}
\end{equation}
\begin{equation}\label{UB}
U_{B}=\int dxdy\frac{Yb^3}{24(1-\nu^2)}\left\{
\left[ {\rm Tr}(H)\right]^2-2(1-\nu){\rm det}(H)\right\}
\end{equation}
where $H$ is the Hessian matrix
\begin{equation}
H(x,y)=\left(
\begin{array}{cc}
\frac{\partial^2 w}{\partial x^2} & \frac{\partial^2 w}{\partial x\partial y} \\
\frac{\partial^2 w}{\partial x\partial y} & \frac{\partial^2 w}{\partial y^2}
\end{array}\right).
\end{equation}

In addition to this, we note that the maximum tensile strain experienced by the material is 
given by
\begin{equation}
e_{\max}=\max_{\theta}\left[
\hat{\bf q}^T{\bf e}\hat{\bf q}
+\frac{b}{2}
\left|
\hat{\bf q}^T
\left(
\begin{array}{cc}
\partial_x^2 w & \partial _{xy}^2 w \\
\partial_{xy}^2 w & \partial _y^2 w 
\end{array}
\right)
\hat{\bf q}
\right|
\right],
\end{equation}
where $\hat{\bf q}^T=(\cos\theta,\sin\theta)$. 
In this expression, the first term comes from stretching from in- and 
out-of-plane deformation of the middle plane
and the second from stretching produced by bending of the plate (which
is maximal at a distance $\pm b/2$ from the middle plane).

In what follows, we assume that the material is brittle, 
in that it will fail if it experiences a 
tensile strain that exceeds a small number $\epsilon\ll 1$. We therefore have the no-breaking 
condition: 
\begin{equation}e_{\max}\le\epsilon\label{no_break}
\end{equation}
We can now state precisely the problem of finding a mechanically efficient 
plate: for a fixed aperture geometry, and for each pair of values 
$(\epsilon,p)$, we wish to find that function $b(x,y)$ which minimises the 
integral $\int{b(x,y)}dxdy$, while at the same time achieving mechanical 
equilibrium (minimising the energy of Eq.~(\ref{U})) and subject to the 
condition of no breakage in Eq.~(\ref{no_break}). In fact, this problem as 
stated is not well-posed; we therefore choose the extra restriction (discussed 
in section~\ref{sparsec}) that $|\nabla b|\le G$ everywhere, for some fixed 
number $G$.

It is tempting to approach this problem by choosing an arbitrary 
function $b(x,y)$, minimising the energy, and then locally removing or 
adding material according as whether $e_{\max}$ is greater than or less 
than $\epsilon$. This is a possible strategy, but based on the calculations 
to follow we suspect that starting from a uniform $b$, this strategy will only 
produce a local minimum in the amount of material used.

\section{One dimensional case}
\subsection{Solution for cantilever boundary conditions}
Let us consider the case where we have a plate which is infinitely long in 
the $y$-direction. Furthermore, we impose the strong restriction that $b$ 
is a function only of $x$, the direction across the plate. This will turn out 
to be very sub-optimal, but it is an instructive case for developing the 
argument. Under these conditions, $v=0$, while $u$, $w$ and $b$ are functions 
only of $x$. The energy per unit length in the $y$-direction of the plate 
under pressure becomes
\begin{eqnarray}\label{U1D}
U=\frac{Y}{1-\nu^2}\int{dx}
\left[
\frac{b}{2}(\partial_x u)^2+
\frac{b}{2}(\partial_x u)(\partial_x w)^2
\right.
\nonumber \\
\left.
\frac{b}{8}(\partial_x w)^4+
\frac{b^3}{24}(\partial_x^2 w)^2
-(1-\nu^2)pw
\right],
\end{eqnarray}
which must be minimised subject to the no-breakage condition 
\begin{equation}
\frac{\partial u}{\partial x}+\frac{1}{2}\left(
\frac{\partial w}{\partial x}\right)^2
+\frac{b}{2}\left|\frac{\partial^2 w}{\partial x^2}\right|
\le\epsilon .
\end{equation}

As stated above, the material is brittle in the sense $\epsilon\ll 1$, and we 
want the plate to be thin ($b/L_0\ll 1$). We also 
expect $u/L_0\ll 1$, $w/L_0\ll 1$, and from the geometry, $O(u)=O(w^2)$, 
which we write for brevity as $u\sim w^2$.

Now, consider first the case where $b$ is a constant and the plate is a strip 
given by $x\in(-a,a)$. Let $\xi\equiv x/a$ then the Euler-Lagrange equations 
obtained by minimising the energy of Eq.~(\ref{U1D}) are
\begin{equation}\label{EL1}
\frac{\partial u}{\partial \xi}=a\frac{b^2}{a^2}\frac{\zeta^2}{12}-
\frac{1}{2a}\left(\frac{\partial w}{\partial\xi}\right)^2 ,
\end{equation}
\begin{equation}\label{EL2}
\frac{\partial^4 w}{\partial\xi^4}=\zeta^2\frac{\partial^2 w}{\partial\xi^2}
+12a\left(\frac{a}{b}\right)^3 \tilde{p} ,
\end{equation} 
where $\tilde{p}\equiv (1-\nu^2)p$ and $\zeta$ is the constant of 
integration of the Euler-Lagrange equations
which is still to be determined (physically,
$\zeta$ has the interpretation that $\zeta^2 b^2/(12a^2)$ is the 
strain experienced by the mid-plane of the plate). 

Noting that $w=\partial_x w=0$ at 
$x=\pm a$, and that $w$ is symmetric about $x=0$, we find 
\begin{equation}
w=\frac{6a\tilde{p}}{\zeta^2}\left(\frac{a}{b}\right)^3
\left[(1-\xi^2)+\frac{2\left[\cosh(\zeta\xi)-\cosh\zeta\right]}{\zeta\sinh\zeta}\right]
\end{equation}
\begin{eqnarray}
u=a\frac{b^2\zeta^2}{12a^2}\xi-\frac{72a\tilde{p}^2}{\zeta^4}\left(\frac{a}{b}\right)^6
\left[\frac{\xi^3}{3}-\frac{2\xi\cosh(\zeta\xi)}{\zeta\sinh\zeta}
\right.
\nonumber \\ 
\left.
+\frac{2\sinh(\zeta\xi)}{\zeta^2\sinh\zeta}+
\frac{\sinh(2\zeta\xi)-2\zeta\xi}{4\zeta\sinh^2\zeta}
\right].
\end{eqnarray}
The condition which determines $\zeta$ is that $u=0$ at $x=\pm a$, which gives
\begin{equation}
1=\frac{864\tilde{p}^2}{\zeta^6}\left(\frac{a}{b}\right)^8
\left[\frac{1}{3}-\frac{2\cosh\zeta}{\zeta\sinh\zeta}+\frac{2}{\zeta^2}
+\frac{\sinh(2\zeta)-2\zeta}{4\zeta\sinh^2\zeta}
\right].
\end{equation}
The Euler-Lagrange equation, Eq.~(\ref{EL1}) for $u$ shows that the 
strain produced by stretching at the middle plane
is uniform over the plate, and therefore the no-breakage condition is
\begin{equation}\label{no_break_1d}
\frac{\zeta^2 b^2}{12a^2}+\frac{b}{2a^2}
\left|\frac{\partial^2 w}{\partial\xi^2}\right|
\le \epsilon.
\end{equation}
 
\subsubsection{Breakage dominated by stretching: $\zeta$ large}
Consider first the case $\zeta$ large, which corresponds to  
$\tilde{p}a^4/b^4\gg 1$, then bending occurs mainly in 
thin ``boundary layers'' near the edges of the plate (i.e. close to 
$\xi=1$ and $\xi=-1$) which have a width $\Delta\xi\sim 1/\zeta$. 
In this limit, we find
\begin{equation}
\zeta\approx(288)^{1/6}\tilde{p}^{1/3}\left(\frac{a}{b}\right)^{4/3}.
\end{equation} 
From the no-breakage condition of Eq.~(\ref{no_break_1d}), we can calculate 
the required thickness of the plate. We find that bending and stretching 
both contribute significantly in the boundary layer (but bending is 
unimportant elsewhere), and the necessary minimum thickness $b_{\min}$ is
\begin{equation}\label{zlcb}
\left(\frac{b_{\min}}{a}\right)\approx 2.885\tilde{p}\epsilon^{-3/2}.
\end{equation}
It is also interesting to know $\zeta$ in terms of $\tilde{p}$ 
and $\epsilon$ for the case when $b=b_{\min}$, in order to estimate 
the thickness of the boundary layers. The result is
\[\zeta\approx 0.626\epsilon^2 \tilde{p}^{-1}.
\]
Lastly, consistency with the fact that $\tilde{p}a^4/b^4\gg 1$ implies 
that $\tilde{p}\ll\epsilon^2$.

\subsubsection{Breakage dominated by bending: $\zeta$ small}
Next, let us consider the opposite extreme, where breakage is 
dominated by bending everywhere; i.e. $\zeta$ is very small, 
equivalent to $\tilde{p}a^4/b^4\ll 1$. This leads to
\[
\zeta\approx\frac{8}{\sqrt{35}}\tilde{p}\left(\frac{a}{b}\right)^4 .
\]
There are no boundary layers, and the expression for $w$ becomes
\[
w\approx\frac{a\tilde{p}}{2}\left(\frac{a}{b}\right)^3\left(
1-\xi^2 \right)^2 .
\]
 
The no-breakage condition is therefore:
\[
\frac{\zeta^2 b^2}{12a^2}+\frac{2\tilde{p}a^2}{b^2}\le\epsilon,
\]
where the first term corresponds to strain from stretching of the middle layer
(both in-plane and out-of-plane), and the second term is strain from bending. 
We find that breakage is governed only by the bending term in this limit, and
\[
\left(\frac{b_{\min}}{a}\right)\approx 2^{1/2}\tilde{p}^{1/2}\epsilon^{-1/2}.
\]
Consistency with $\tilde{p}a^4/b^4\ll 1$ leads to $\tilde{p}\gg\epsilon^2$, 
and thinness of the plate 
$(b/a)\ll 1$ also imposes the condition $\tilde{p}\ll\epsilon$.

\subsection{Freely hinged boundary conditions}
In the case of $\zeta$ large, the required minimum thickness of a uniform plate is determined by 
the thickness needed to sustain the bending and stretching forces in the thin boundary 
layers at the edges of the plate. Given more freedom to design a plate with non-uniform 
thickness, one might imagine that the most efficient design would have a 
larger value for $b$ in the two boundary layers near the edges, and be 
thinner everywhere else. The required thickness of this middle region 
could be estimated conservatively from the minimum thickness required for 
a plate with freely hinged boundary conditions. This case is also 
immediately solvable: we require that $w$ be symmetrical while 
$u=0$, $w=0$ and $\partial_x^2 w=0$ at 
$x=\pm a$. The Euler-Lagrange equations Eq.~(\ref{EL1}) and (\ref{EL2}) then
give in the limit of large $\zeta$
\begin{equation}\label{fh}
\left(\frac{b_{\min}}{a}\right)\approx
0.408\tilde{p}\epsilon^{-3/2}.
\end{equation}
The important thing to note is that although the pre-factor is smaller in Eq.~(\ref{fh}), 
the order of 
magnitude of all quantities is the same as for the cantilevered boundary condition 
case (Eq.~(\ref{zlcb})). 
Even if we were to optimise the mechanical efficiency of the plate by 
choosing $b$ to be thicker in the boundary layer and thinner in the centre of 
the plate, we would only achieve a gain in efficiency by a numerical factor 
of order unity. We also conjecture that an optimisation strategy comprising 
removing material where the breakage condition is not satisfied will 
converge on a solution of this kind. 

The boundary layers we encounter in this case are therefore in some sense 
unimportant; having only a quantitative and not a qualitative effect on the 
amount of material needed to make our pressure-bearing plate. Shortly 
however, we shall encounter cases where the presence of boundary layers 
(regions where quantities vary rapidly in space) 
alters even the scaling of mechanical efficiency with $p$ and $\epsilon$.

\subsection{Order of magnitude behaviour for a uniform plate}
We can now state the order-of-magnitude behaviour of various quantities 
(assuming $\nu\neq -1$) for our optimum 
thin plate which has either uniform thickness, or boundary layers at the edges. In what follows,
we shall use ``$\sim$'' to denote that two quantities are the same order of magnitude. We have:
\begin{equation}\label{b_uniform}
\left(\frac{b_{\min}}{a}\right)\sim
\left\{
\begin{array}{cc}
p\epsilon^{-3/2} & p\ll\epsilon^2 \\ 
p^{1/2}\epsilon^{-1/2} & \epsilon^2\ll p\ll\epsilon
\end{array}
\right. ,
\end{equation}
where the first case corresponds to a stretching-dominated and the second to a 
bending-dominated regime. 

\begin{figure}
\includegraphics[width=\columnwidth]{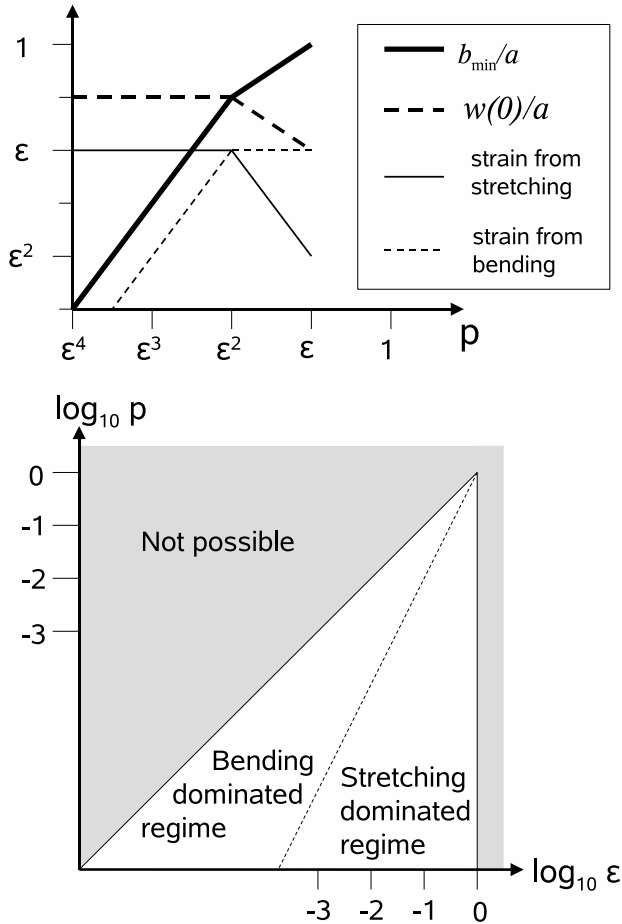}
\caption{\label{oom1} 
Top figure: order of magnitude behaviour for quantities as a function of pressure, for an
aperture spanned by a plate of uniform thickness. Plotted on this 
scale, all crossovers are sharp and all order 1 prefactors are invisible. Bottom figure: 
schematic picture of the $p-\epsilon$ plane, showing the approximate positions of crossovers
in behaviour for this case. The dashed line showing the crossover of regimes is $p\sim\epsilon^2$}
\end{figure}

The maximum deflection, maximum curvature from bending and maximum strain from 
stretching (both in-plane and out of plane) are given by
\begin{equation}
\left(\frac{w(0)}{a}\right)\sim
\left\{
\begin{array}{cc}
\epsilon^{1/2} & p\ll\epsilon^2 \\ 
p^{-1/2}\epsilon^{3/2} & \epsilon^2\ll p\ll\epsilon
\end{array}
\right. ,
\end{equation}
\begin{equation}
b\left.\frac{\partial^2 w}{\partial x^2}\right|_{\max}\sim
\left\{
\begin{array}{ccc}
\epsilon & p\ll\epsilon^2 & {\rm in\ b.\ layer} \\ 
p\epsilon^{-1} & p\ll\epsilon^2 & {\rm out\ b.\ layer} \\ 
\epsilon & \epsilon^2\ll p\ll\epsilon & \ 
\end{array}
\right. ,
\end{equation}
\begin{equation}
\frac{\zeta^2 b^2}{a^2}\sim
\left\{
\begin{array}{cc}
\epsilon & p\ll\epsilon^2 \\ 
p^{-1}\epsilon^{3} & \epsilon^2\ll p\ll\epsilon
\end{array}
\right. .
\end{equation}
These relations are shown in the top panel of Fig.~\ref{oom1} 
as an ``order of magnitude'' (``o.o.m.'') plot. The
advantage of the compressed scale used in this plot is that all crossover behaviour is sharp,
and prefactors are irrelevant (indeed even a factor as large as $\log(\epsilon^{-1})$ would
be invisible on this scale). The bottom panel of Fig.~\ref{oom1} shows the different
regimes in the $p-\epsilon$ plane. Because of the different scale, pre-factors are now important,
and so the picture must be regarded as schematic (except very far from the origin at
$(\log\epsilon,\log p)=(0,0)$ ).

\section{Plate with parallel spars}
\subsection{Behaviour of a single spar\label{sparsec}}
\begin{figure}
\includegraphics[width=\columnwidth]{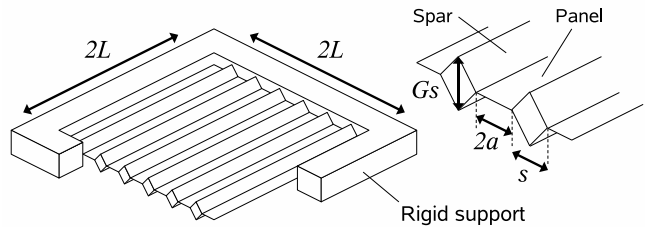}
\caption{\label{parallel_spars} 
Schematic for the design of a plate consisting of parallel spars separated by long narrow panels, 
and spanning a square aperture of side $2L$.}
\end{figure}

So far we have analysed the case where the thickness $b$ is constant, or (by arguing from 
these solutions), the case where there are narrow boundary layers at the edges of the plate 
accommodating the local bending forces. However, because the bending term (Eq.~(\ref{UB})) 
in the energy 
(Eq.~(\ref{U})) has a pre-factor cubic in the 
plate thickness, we expect intuitively that it would be more 
efficient still to design a plate with narrow, thick (large $b$) beams or ``spars'', which support 
thinner ``panels'' between. A simple example of such a design would be as shown in 
Fig.~\ref{parallel_spars}.

Again, from the form of the energy expression, we would expect that the design would get 
more efficient the narrower (smaller $s$ in Fig.~\ref{parallel_spars}) and 
the taller (larger $Gs$ in Fig.~\ref{parallel_spars}) the 
supporting spars are designed to be. However, at some point, very narrow and tall spars are 
likely to buckle under loading. In the analysis that follows, we have therefore placed a simple 
restriction on the design of the plate, which is both interesting in its own right, but can also 
be seen as an attempt to capture this buckling limitation. The design restriction we place is 
that the maximum  gradient of the thickness cannot exceed some set value $G$,
which is of order unity.
There is also a third reason for choosing such a $G$, which represents a 
limitation
of the current analysis, rather than a change in the qualitative beahviour
of the system. This limitation is that if the thickness $b$ varies
too quickly in space, then Eqs.~(\ref{US}) and (\ref{UB}) will no longer be
valid.

Stated formally the new restriction is that
\begin{equation}
\forall(x,y):\left|\nabla b(x,y)\right|\le G.
\end{equation}
Let us therefore consider a spar which has a diamond-shaped cross section 
satisfying this constraint, with width $s$, length $2L$ and loaded under a 
force $\tilde{f}$  per unit length.  We define a re-scaled parameter with 
dimensions of length to describe this loading, namely $f=\tilde{f}/Y$. 

In order to write down the energy of this system from Eq.~(\ref{U}), we need
a further condition on the deformation of the middle plane. Because
we are envisaging a narrow spar, we take this condition to be that the
component of the stress in the $x$-direction (across the spar) at the middle 
plane is zero.
Because the $x$ and $y$ directions are still the principal directions
for both the strain and stress tensor, then this leads to $e_{xx}=-\nu e_{yy}$
and so the energy of the spar is given by
\begin{eqnarray}\label{Ubeam}
U=Y\int{dy}\left\{
\frac{Gs^2}{4}\left[\partial_y v+\frac{1}{2}(\partial_y w)^2\right]^2
\right.
\nonumber \\
\left.
+\frac{G^3 s^4}{192}\frac{(\partial_y^2 w)^2}{(1-\nu^2)}-fw
\right\}.
\end{eqnarray}
Defining $\eta\equiv y/L$, the Euler-Lagrange equations for Eq. (\ref{Ubeam}) 
are 
\[
\frac{\partial v}{\partial\eta}=L\frac{\zeta^2G^2s^2}{48L^2(1-\nu^2)}
-\frac{1}{2L}\left(\frac{\partial w}{\partial\eta}\right)^2 ,
\]
\[
\frac{G^2 s^2}{48 L^2}\frac{\partial^4 w}{\partial\eta^4}=
\frac{\zeta^2 G^2 s^2}{48L^2}\frac{\partial^2 w}{\partial\eta^2}
+\frac{2f(1-\nu^2)L^2}{Gs^2} ,
\]
where $\zeta$ is a constant of integration in the Euler-Lagrange equations for
Eq.~(\ref{Ubeam}) and needs to be determined. Solving these with 
cantilever boundary conditions gives
\[
w=\frac{48L^4 (1-\nu^2)f}{G^3 s^4 \zeta^2}
\left[ (1-\eta^2)+\frac{2\left[\cosh(\zeta\eta)-\cosh\zeta\right]}{\zeta\sinh\zeta}
\right],
\]
\begin{eqnarray}
v=\frac{\zeta^2 G^2 s^2}{48L(1-\nu^2)}\eta
\nonumber \\
-\frac{4608L^7 (1-\nu^2)^2 f^2}{G^6 s^8 \zeta^4}
\left[\frac{\eta^3}{3}-\frac{2\eta\cosh(\zeta\eta)}{\zeta\sinh\zeta}
\right.
\nonumber \\ 
\left.
+\frac{2\sinh(\zeta\eta)}{\zeta^2\sinh\zeta}
+\frac{\sinh(2\zeta\eta)-2\zeta\eta}{4\zeta\sinh^2\zeta}
\right],
\nonumber
\end{eqnarray} 
where the constant of integration $\zeta$ can be determined from
\begin{eqnarray}
\frac{G^8 s^{10}}{L^8 f^2}=
\frac{221184(1-\nu^2)^3}{\zeta^6}
\nonumber \\
\times\left[\frac{1}{3}-\frac{2\cosh\zeta}{\zeta\sinh\zeta}+\frac{2}{\zeta^2}
+\frac{\sinh(2\zeta)-2\zeta}{4\zeta\sinh^2\zeta}
\right].\nonumber
\end{eqnarray}
From now on, we consider only the orders of magnitudes of different 
quantities, and again there will be two cases: 

\subsubsection{Stretching dominated regime}
In the stretching dominated regime, we have $\zeta\gg 1$, and
\[
\zeta\sim\frac{L^{4/3}f^{1/3}}{G^{4/3}s^{5/3}},
\]
so $f\gg G^4 s^5/L^4$  and the no-breaking condition (which is dominated by stre
tching
everywhere except at the ends of the spar, where bending is also significant) gi
ves
\[
f\le \epsilon^{3/2}s^2 L^{-1}G .
\]
Because of the no-breakage condition, the strain from in-plane and out-of-plane
stretching is
always less than $\epsilon$. The maximum amount of bending is
\[
\left.\frac{\partial^2 w}{\partial y^2}\right|_{\max}\sim
\frac{L^2 f}{G^3 s^4 \zeta}\sim\frac{L^{2/3} f^{2/3}}{G^{5/3}s^{7/3}} ,
\]
which occurs in the ``boundary layers'' near the two ends of the spar.
The maximum out-of-plane deflection is
\[w(0)\sim\frac{L^{4/3}f^{1/3}}{G^{1/3}s^{2/3}}.\]

\subsubsection{Bending dominated regime}
In the case that $\zeta$ is small, we are in the bending dominated regime and we find that
\[
\zeta\sim\frac{L^4 f}{G^4 s^5}
\]
Therefore consistency requires that $f\ll G^4 s^5/L^4$  and the no-breaking condition (which is 
dominated by bending) is
\[
f\le\epsilon s^3 L^{-2}G^2 .
\]
The maximum amount of bending is given by
\[ 
\left.\frac{\partial^2 w}{\partial y^2}\right|_{\max}\sim
\frac{L^2 f}{G^3 s^4}
\]
and the amount of in-plane and out-of-plane stretching is of order
\[
\frac{\zeta^2 G^2 s^2}{L^2}\sim\frac{L^6 f^2}{G^6 s^8}.
\] 
The maximum out-of plane deflection is
\[w(0)\sim\frac{L^4 f}{G^3 s^4}. \] 

\subsection{Parallel spars supporting panels}
Let us suppose we have a plate consisting, as in Fig.~\ref{parallel_spars}, of 
parallel ``spars'' separated 
by thinner ``panels''. The parallel spars also represent a collection of boundary layers for the 
function $b(x,y)$, in the sense that quantities have a fast variation as a function of $x$ or $y$. In 
contrast to the case discussed above, these boundary layers produce a qualitative change in 
the system's behaviour. 

We assume that the entire aperture is a square of side $2L$, and that the spars are of width $s$, 
and separated by a distance $2a$ from neighbouring spars. Because the analysis from now on 
will focus on orders of magnitude, we shall assume that $G$ is of order unity, and omit it from 
the expressions. We will assume (and later check) that $s\ll a$ and $a\ll L$ (the latter, together 
with the relative rigidity of the spars, ensuring that we can treat the panels as infinitely long 
strips). The panels have a thickness $b_0$.

We now calculate the average thickness $\langle b\rangle$ of the plate that 
is required to support a pressure load, where (again in order-of-magnitude 
terms)
\begin{equation}\label{bangle1}
\langle b\rangle\sim
\frac{ab_0+s^2}{a+s}\sim\frac{\max(ab_0,s^2)}{\max(a,s)},
\end{equation}
and the second equivalence follows from the fact that the terms being added (in
each of the numerator and denominator) will in general be of different 
magnitude.

Because $s\ll a$, then over almost all their length, the spars must support a force per unit 
length given by $f=2ap$.
 
We consider the four cases where the spars and panels are in the stretching and bending 
dominated regimes. There are six conditions which have to be applied in each case:

(i) The panels are assumed to be at their most mechanically efficient, which sets $b_0$ as a 
function of $p$ and $a$.

(ii) Whether the panels are in the bending or stretching dominated regime sets a constraint 
on the pressure.

(iii) The spars are in the stretching dominated regime if $ap\gg s^5/L^4$  and the bending 
dominated regime if $ap\ll s^5/L^4$.

(iv) The spars must not break (although we do not insist that they should be at this limit; they 
may be somewhat over-engineered in this sense). 

(v) The spars must provide significant support, in the sense that the maximum out-of-plane 
displacement at the centre of each spar must be much less than the maximum out-of-plane 
displacement at the centre of each panel.

(vi) There must be at least one spar in the system; in other words $(a/L)\le 1$.

These conditions are listed in Table~\ref{conditions}.

\begin{table}
\caption{\label{conditions} The six conditions which must be satisfied by spars and panels
for a plate consisting of parallel spars supporting intervening panels. Spars and panels may be 
either in the bending or stretching dominated regimes.}
\begin{ruledtabular}
\begin{tabular}{lcc}
\ & Spars in stretching & Spars in bending \\
\ & dominated regime & dominated regime \\
\hline
\          & $b_0\sim ap\epsilon^{-3/2}$ & $b_0\sim ap\epsilon^{-3/2}$ \\
Panels in  & $p\ll \epsilon^2$ & $p\ll \epsilon^2$ \\
stretching & $\left(\frac{s}{L}\right)\ll\left(\frac{a}{L}\right)^{1/5}p^{1/5}$ & $\left(\frac{s}{L}\right)\gg\left(\frac{a}{L}\right)^{1/5}p^{1/5}$ \\
dominated  & $\left(\frac{s}{L}\right)\ge\left(\frac{a}{L}\right)^{1/2}p^{1/2}\epsilon^{-3/4}$ & $\left(\frac{s}{L}\right)\ge\left(\frac{a}{L}\right)^{1/3}p^{1/3}\epsilon^{-1/3}$ \\
regime     & $\left(\frac{s}{L}\right)\gg\left(\frac{a}{L}\right)^{-1}p^{1/2}\epsilon^{-3/4}$ & $\left(\frac{s}{L}\right)\gg p^{1/4}\epsilon^{-1/8}$ \\
\          & $(a/L)\le 1$ & $(a/L)\le 1$ \\
\hline
\          & $b_0\sim ap^{1/2}\epsilon^{-1/2}$ & $b_0\sim ap^{1/2}\epsilon^{-1/2}$ \\
Panels in  & $\epsilon^2\ll p\ll\epsilon$ & $\epsilon^2\ll p\ll\epsilon$ \\
bending    & $\left(\frac{s}{L}\right)\ll\left(\frac{a}{L}\right)^{1/5}p^{1/5}$ & $\left(\frac{s}{L}\right)\gg\left(\frac{a}{L}\right)^{1/5}p^{1/5}$ \\
dominated  & $\left(\frac{s}{L}\right)\ge\left(\frac{a}{L}\right)^{1/2}p^{1/2}\epsilon^{-3/4}$ & $\left(\frac{s}{L}\right)\ge\left(\frac{a}{L}\right)^{1/3}p^{1/3}\epsilon^{-1/3}$ \\
regime     & $\left(\frac{s}{L}\right)\gg\left(\frac{a}{L}\right)^{-1}p^{5/4}\epsilon^{-9/4}$ & $\left(\frac{s}{L}\right)\gg p^{3/8}\epsilon^{-3/8}$ \\
\          & $(a/L)\le 1$ & $(a/L)\le 1$ \\
\end{tabular}
\end{ruledtabular}
\end{table}

\begin{table}
\caption{\label{props_of_soln} The scaling properties of optimally
efficient plates consisting of parallel spars supporting intervening panels. 
Spars and panels may be 
either in the bending or stretching dominated regimes.}
\begin{ruledtabular}
\begin{tabular}{lcc}
\ & Spars in stretching & Spars in bending \\
\ & dominated regime & dominated regime \\
\hline
Panels in  & $p\ll \epsilon^{5/2}$ & $\epsilon^{5/2}\ll p\ll \epsilon^2$ \\
stretching & Spars give & $\langle b\rangle\sim L p^{3/4}\epsilon^{-7/8}$ \\
dominated  & no gain & $\left(\frac{s}{L}\right)\sim p^{1/4}\epsilon^{-1/8}$ \\
regime     & in efficiency & $\left(\frac{a}{L}\right)\sim p^{-1/4}\epsilon^{5/8}$ \\
\hline
Panels in  & Not & $\epsilon^2\ll p\ll\epsilon$ \\
bending    & consistent & $\langle b\rangle\sim L p^{5/8}\epsilon^{-5/8}$ \\
dominated  & with & $\left(\frac{s}{L}\right)\sim p^{3/8}\epsilon^{-3/8}$ \\
regime     & conditions & $\left(\frac{a}{L}\right)\sim p^{1/8}\epsilon^{-1/8}$ \\
\end{tabular}
\end{ruledtabular}
\end{table}

It is a simple, but tedious matter to combine the conditions of 
Table~\ref{conditions} with Eq.~(\ref{bangle1}) to obtain
the scaling properties of the optimally efficient solution in each case. 
The results are shown in Table~\ref{props_of_soln}.

\begin{figure}
\includegraphics[width=\columnwidth]{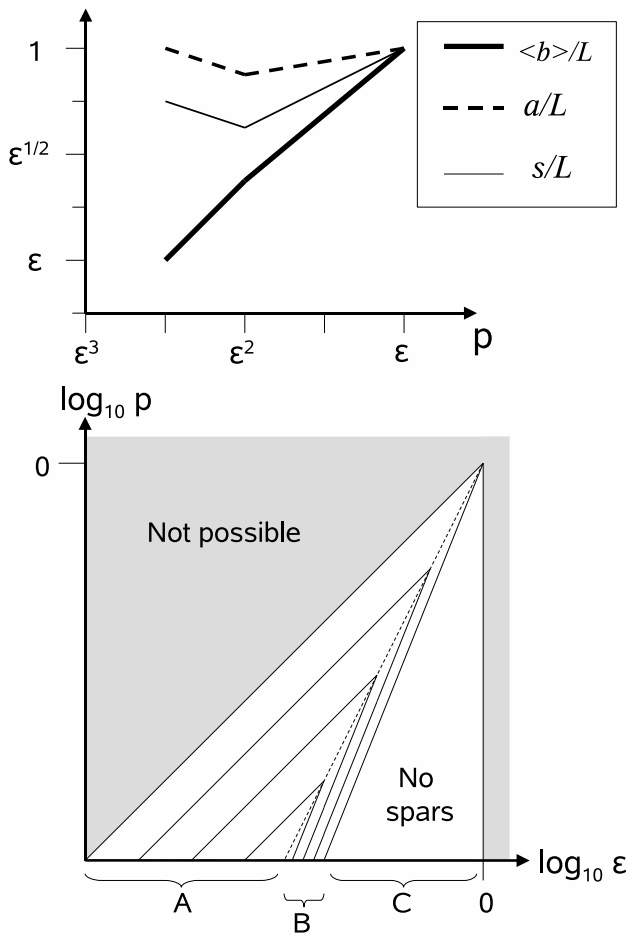}
\caption{\label{oom2}
Top figure: order of magnitude behaviour for quantities as a function of 
pressure for an
aperture spanned by a plate consisting of parallel spars supporting long panels.
Bottom figure:
schematic picture of the $p-\epsilon$ plane, showing contours (solid lines) 
of constant number of
spars in the structure. This number increases towards the bottom left of the 
figure. Region
`A' is where spars and panels are bending dominated 
($\epsilon^2\ll p\ll\epsilon$). Region `B'
is where spars are bending dominated and panels are stretching dominated
($\epsilon^{5/2}\ll p\ll\epsilon^2$), and region `C' is where
there are no spars, and the now uniformly
thick plate is stretching dominated ($p\ll\epsilon^{5/2}$).}
\end{figure}

Fig.~\ref{oom2} shows these behaviours on an o.o.m. plot, and (schematically) 
on the $p-\epsilon$
plane.

\section{Fractal design}
\subsection{Two beams at right angles}
The parallel spars analysed above have provided considerable gains in efficiency over a 
uniform plate, but it is possible that we have not achieved the global optimal scaling. One 
possibility which remains is that we could allow spars to intersect. In this way, thicker spars 
may be used to partially support thinner spars, which together support panels. We can 
envisage ultimately a hierarchical arrangement with spars of progressively thinner aspect 
forming a fractal design, and supporting panels only at the smallest length scales.

To proceed, we first investigate whether the principle of a thicker spar 
supporting a thinner spar can indeed lead to a gain in efficiency for a 
structure. Consider therefore a single spar which is in the bending dominated 
regime, carrying a (scaled) force $f$ per unit length, and a point 
load $\tilde{F}$  at the centre, which we represent by a scaled point 
force $F=\tilde{F}/Y$. The spar or beam must be considered in two sections, 
joined at the middle ($x=0$), at which point the 
displacement $w$ and the first two derivatives of $w$ are continuous
\begin{equation}
w(x)=\left\{
\begin{array}{cc}
w^{+}(x) & x\in(0,L) \\
w^{-}(x) & x\in (-L,0)
\end{array}
\right.
\end{equation}

Hence, if we have two beams at right angles with widths $s_1$ and $s_2$, joined at the centre, and 
each carrying a force $f$ per unit length, they will (in the absence of external point forces) 
apply equal and opposite forces $F$ on each other, and their displacements will be given by the 
following expressions:
\[
w_{1}^{\pm}(x)=\frac{4f}{G^3 s_{1}^{4}}
\left(x^2-L^2\right)^2\pm
\frac{8F}{G^3 s_1^4}\left(x\mp L\right)^2 \left(x\pm \frac{L}{2}\right)
\]
\[
w_{2}^{\pm}(y)=\frac{4f}{G^3 s_{2}^{4}}
\left(y^2-L^2\right)^2\mp
\frac{8F}{G^3 s_2^4}\left(y\mp L\right)^2 \left(y\pm \frac{L}{2}\right)
\]
\begin{equation}\label{F}
F=fL\left(\frac{s_1^4-s_2^4}{s_1^4+s_2^4}\right),
\end{equation}
where the point force $F$ in Eq.~(\ref{F}) imposes the constraint that the beams must have the 
same displacement $w(0)$ at their centre points.

   For this highly symmetrical case, the maxima in curvature occur either at the ends or the 
centres of the beams (for more general cases, there may be maxima between the positions 
of application of point loads). We can therefore write the no-breakage condition on the first 
beam as
\begin{eqnarray}
\frac{Gs_1}{2}\left|\partial_x^2 w_1^\pm (x)\right|_{\max}=
\nonumber \\
\frac{4fL^2}{G^2 s_1^3 (r^4+1)}
\max\left[ \left|5r^4-1\right|,(7r^4+1)\right]\le\epsilon
\nonumber
\end{eqnarray}
or
\begin{equation}\label{s1}
\frac{s_1^3}{L^3}\ge
\frac{4f}{(r^4+1)LG^2\epsilon}\max\left[ \left|5r^4-1\right|,(7r^4+1)\right],
\end{equation}
where $r\equiv s_1/s_2$. The result must be symmetric on swapping the two spars, so
\begin{equation}\label{s2}
\frac{s_2^3}{L^3}\ge
\frac{4f}{(r^{-4}+1)LG^2\epsilon}\max\left[ \left|5r^{-4}-1\right|,(7r^{-4}+1)\right].
\end{equation}
 
The total volume of material used to make these beams is therefore
\begin{equation}
V=GL\left(s_1^2+s_2^2\right),
\end{equation} 
which we seek to minimise.

   Eqs.~(\ref{s1}) and (\ref{s2}) have a trivial solution in which $s_1=s_2$ and $F=0$. The 
most efficient solution of this kind has both beams at the breaking limit, so 
\begin{equation}
s_1=s_2=2^{4/3}L(f/L)^{1/3}\epsilon^{-1/3}G^{-2/3},
\end{equation}
and uses a volume of material given by
\begin{equation}
V\approx 12.699 L^3 (f/L)^{2/3}\epsilon^{-2/3}G^{-1/3}.
\end{equation} 
However, this is not the global optimum, which in fact occurs when $r\approx 1.22$ and
\begin{equation}
V\approx 12.652 L^3 (f/L)^{2/3}\epsilon^{-2/3}G^{-1/3},
\end{equation} 
and only the thinner of the two beams is at its breaking limit.

   We therefore see that although the gains are very modest, it is nevertheless possible to 
achieve greater efficiency through having a thicker beam or spar support a thinner. The next 
question to address is whether this principle can be extended to a hierarchical arrangement 
with thin spars supporting still thinner spars, until the very thinnest help to support panels to 
form a continuous plate with no holes.

\subsection{Hierarchical arrangement of spars}
\begin{figure}
\includegraphics[width=\columnwidth]{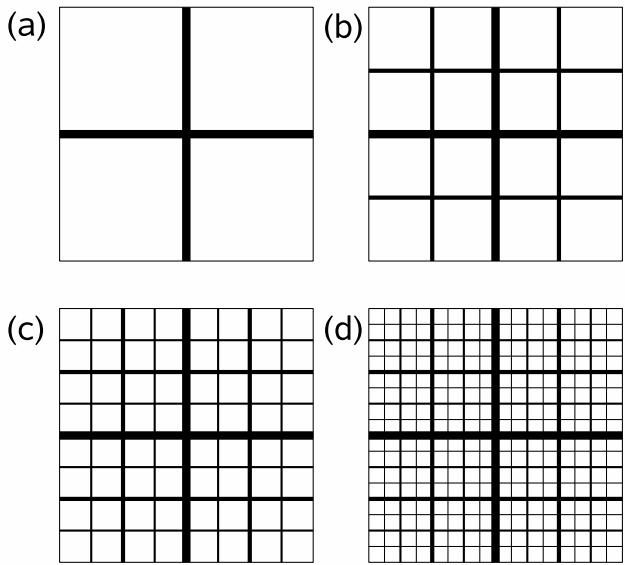}
\caption{\label{fractal_design} 
Schematic of a fractal design for a plate spanning a square aperture. (a) is a ``generation 1''
structure, with two crossed spars supporting four panels. In (b), four thinner spars are
added, which subdivides the remainder of the plate into sixteen panels. In (c) and (d) we show
generations 3 and 4.}
\end{figure}

Consider the arrangement of spars as shown in Fig.~\ref{fractal_design}. 
Fig.~\ref{fractal_design}(a) shows two spars at right 
angles, and we refer to this structure as ``generation 1''. We choose the spars to have the 
same width $s_{1,1}$ (the first index referring to the ``generation number'', and the second to the 
type of spar present in this generation). Each spar carries a force per unit length $f_1=f_0/2$. In 
light of the previous calculation, we know that it is not optimal to have both spars of the same 
width; but we ignore this fact in favour of simplicity of analysis. 

   Fig~\ref{fractal_design}(b) shows a ``generation 2'' structure, 
where we have two thick spars, just as for 
generation 1, but their widths are now $s_{2,1}$ which may be different from $s_{1,1}$. 
In addition, there 
are four thinner spars, each with width $s_{2,2}$. All the spars now carry a force per unit length of 
$f_2=f_0/4$ --  chosen to approximate the loading produced by the smaller panels between the 
spars which will eventually be present in the structure. Fig.~\ref{fractal_design}(c) and (d) 
show generations 3 
and 4 respectively, and we envisage going to large generation numbers $n$ for certain regions 
of the $p-\epsilon$ plane. 

   For generation $n$, we will have two spars of width $s_{n,1}$, four thinner spars 
of width $s_{n,2}$ and so on, culminating with $2^n$ of the thinnest spars, which each have a width 
$s_{n,n}$. The total number of spars in the generation $n$ structure is therefore $2(2^n-1)$, 
and each 
carries a force per unit length (in addition to the point forces) of $f_n=2^{-n}f_0$. 
Again, this is 
chosen to mimic the loading on the spars produced by supported panels, so that $f_{0}\approx pL/4$
(although the loading in this final structure will not be exactly uniform).

We now analyse the first few generations numerically, ignoring the possible 
torsional loadings that spars can exert upon one another in asymmetrical 
configurations. We expect this complication to have minimal effects 
when $G\lesssim 1$, and so we now specialise to the case $G=1$.

   For generation $n$, we choose a structure described by the two parameters $\beta (n)$
and $\gamma (n)$ defined by
\begin{equation}\label{snq}
s_{n,q}=L\left(\frac{f_0}{L}\right)^{1/3}\epsilon^{-1/3}\beta (n)
\left[\gamma (n)\right]^{q-1}.
\end{equation}
We choose $\beta (n)$ and $\gamma (n)$ so that all the spars satisfy the no-breaking condition, and 
the amount of material V that is used, is minimised. Adding up the total volume of material in 
the spars gives 
\begin{equation}\label{Vn}
V(n)=2L^3\left(\frac{f_0}{L}\right)^{2/3}
\epsilon^{-2/3}\beta^2(n)\left\{
\frac{\left[2\gamma^2(n)\right]^n-1}{2\gamma^2(n)-1}
\right\}.
\end{equation}
Based on experience from the previous section, we expect that the most efficient structure of 
this kind will satisfy the breakage condition as an equality only for the thinnest spars. We 
also expect $\gamma(n)>0.5$ (the thicker spars will not act as completely rigid 
supports for the thinner spars), but we hope that a hierarchical structure is still achieved, 
which corresponds to $\gamma(n)<1$.

Table~\ref{numerics} shows the first few values for $\beta(n)$ and $\gamma(n)$ which are found from
numerical simulations to minimise $V(n)$. The tentative conclusion is that both quantities
tend to fixed, order 1 values as $n$ increases.

\begin{table}
\caption{\label{numerics} 
Values for $\beta(n)$ and $\gamma(n)$ (in Eq.~(\ref{snq})) which are found numerically to 
minimise $V(n)$ (Eq.~(\ref{Vn}) from the text).}
\begin{ruledtabular}
\begin{tabular}{ccc}
$n$ & $\beta(n)$ & $\gamma(n)$ \\
\hline
1 & 1.260 & $-$ \\
2 & 1.126 & 0.609 \\
3 & 1.009 & 0.644 \\
4 & 1.078 & 0.650
\end{tabular}
\end{ruledtabular}
\end{table}

If now we assume that $\beta(n)$ and $\gamma(n)$ do indeed tend to fixed limits 
$\beta$ and $\gamma$ as $n\rightarrow\infty$, then provided the 
spars remain much narrower than the panels, and the spars remain in the bending 
dominated regime, we can calculate the average thickness of the plate. It will be a sum of 
two terms; one from the hierarchical structure of spars, and one from the panels between 
(which may be in either the stretching or bending dominated regimes). For an order of 
magnitude calculation, the pre-factors are unimportant and we expect square panels to 
exhibit the same scaling as the long thin panels analysed above. By using 
Eqs.~(\ref{Vn}) and (\ref{b_uniform}) we obtain in 
the large $n$ limit, and assuming $\beta$ is an order 1 quantity
\begin{equation}\label{beta_sum}
\frac{\langle b\rangle}{L}\sim\left\{
\begin{array}{cc}
(2\gamma^2)^n p^{2/3}\epsilon^{-2/3} +2^{-n}p\epsilon^{-3/2} & p\ll\epsilon^2 \\
(2\gamma^2)^n p^{2/3}\epsilon^{-2/3} +2^{-n}p^{1/2}\epsilon^{-1/2} & \epsilon^2\ll p\ll\epsilon
\end{array}
\right. .
\end{equation}
 
If we minimise this as a function of n, we find

\begin{equation}\label{2n}
2^n\sim\left\{
\begin{array}{cc}
p^{\frac{\ln 2}{6\ln(2\gamma)}}\epsilon^{-\frac{5\ln 2}{12\ln(2\gamma)}} & p\ll\epsilon^2 \\
p^{-\frac{\ln 2}{12\ln(2\gamma)}}\epsilon^{\frac{\ln 2}{12\ln(2\gamma)}} & \epsilon^2\ll p\ll\epsilon
\end{array}
\right. ,
\end{equation}
\begin{equation}\label{bfracL}
\frac{\langle b\rangle}{L}\sim\left\{
\begin{array}{cc}
p^{\left(1-\frac{\ln 2}{6\ln(2\gamma)}\right)}\epsilon^{\left(-\frac{3}{2}+\frac{5\ln 2}{12\ln(2\gamma)}\right)} & p\ll\epsilon^2 \\
p^{\left(\frac{1}{2}+\frac{\ln 2}{12\ln(2\gamma)}\right)}\epsilon^{\left(-\frac{1}{2}-\frac{\ln 2}{12\ln(2\gamma)}\right)} & \epsilon^2\ll p\ll\epsilon
\end{array}
\right. ,
\end{equation}
where for each of Eqs.~(\ref{2n}) and (\ref{bfracL}), the upper expression corresponds to panels 
in the stretching regime and the lower to panels in the bending regime.

We therefore find that the hierarchical 
arrangement is more efficient than a uniform plate, provided that  
$\epsilon^{5/2}\ll p\ll\epsilon$, but for 
smaller values of p, a plate of uniform thickness is most efficient. This is precisely the range
over which parallel spars are more efficient than uniformly thick plates, so the next question is 
whether a fractal structure is more efficient than the parallel spars? 

By comparing Eq.~(\ref{bfracL}) with the two expressions for $\langle b\rangle$
in Table~\ref{props_of_soln} 
we find that there is a critical value $\gamma_c$ of $\gamma$, above which the parallel 
arrangement is more 
efficient, and below which fractal structures are preferred. The critical value is
\begin{equation}\label{gammac}
\gamma_c=2^{-1/3}\approx 0.7937.
\end{equation}
 
Based on the numerical calculations of Table~\ref{numerics}, it appears that 
$\gamma\approx 0.65<\gamma_c$, and 
so (provided $\epsilon\ll 1$ in the mathematical sense), then the hierarchical structure is 
always more efficient over the entire range $\epsilon^{5/2}\ll p\ll\epsilon$. 

   We note however that for small, but finite values of $\epsilon$, the pre-factors are no longer 
negligible, and it 
remains a possibility that there is a region of the $p-\epsilon$ plane not too far from the origin 
$(\log\epsilon,\log p)=(0,0)$, where a parallel arrangement of spars is preferred.

   Lastly, we note that for the fractal structure, the width of the thinnest spars is given by
\begin{equation}\label{sfracL}
\frac{s_{n,n}}{L}\sim\left\{
\begin{array}{cc}
p^{\left(\frac{1}{3}+\frac{\ln\gamma}{6\ln(2\gamma)}\right)}\epsilon^{\left(-\frac{1}{3}-\frac{5\ln\gamma}{12\ln(2\gamma)}\right)} & p\ll\epsilon^2 \\
p^{\left(\frac{1}{3}-\frac{\ln\gamma}{12\ln(2\gamma)}\right)}\epsilon^{\left(-\frac{1}{3}+\frac{\ln\gamma}{12\ln(2\gamma)}\right)} & \epsilon^2\ll p\ll\epsilon
\end{array}
\right. ,
\end{equation}
so $s_{n,n}\ll a\equiv 2^{-n}L$ for $\epsilon^{5/2}\ll p\ll\epsilon$ and 
$\gamma \in(0.5,1)$, which justifies Eq.~(\ref{beta_sum}) for $\langle b\rangle$.

\section{Conclusions}
We have constructed a simple problem in elasticity theory, in which it is 
possible to frame the question of optimal mechanical efficiency in an easily approachable
manner. We find that it is necessary to consider all regions of the $p-\epsilon$ plane to tackle 
this problem (where $p$ represents the applied pressure and $\epsilon$ the material's brittleness)
and have used order-of-magnitude plots as a concise way to present the 
assymptotic behaviour under the assumption $\epsilon\ll 1$. 

   After considering various possible 
forms for the plate in the problem, we discover regions of the $p-\epsilon$ plane where a 
fractal design is the most mechanically efficient structure we have been able to find. This is a 
structure where thicker spars act as partial support for thinner spars, and so on until the 
thinnest spars (together with all the others) support panels, which form a continuous plate. 

Our analysis has focused on the order of magnitudes of quantities; this greatly simplifies the 
algebra, but leaves open questions about the behaviour for small, but finite values of $p$ and 
$\epsilon$.

\begin{acknowledgements}
I gratefully acknowledge J.J.M. Janssen for inviting me on secondment and providing
a wonderfully supportive environment in which to do research.
\end{acknowledgements}

\end{document}